\documentstyle[12pt,aaspp4]{article}
\def\gtorder{\mathrel{\raise.3ex\hbox{$>$}\mkern-14mu
             \lower0.6ex\hbox{$\sim$}}}
\def\ltorder{\mathrel{\raise.3ex\hbox{$<$}\mkern-14mu
             \lower0.6ex\hbox{$\sim$}}}
\def\ltsima{$\; \buildrel < \over \sim \;$}
\def\simlt{\lower.5ex\hbox{\ltsima}}
\def\gtsima{$\; \buildrel > \over \sim \;$}
\def\simgt{\lower.5ex\hbox{\gtsima}}
\def\rxte{{\it RXTE}}

\begin{document}

\title{A Possible 100-day X-ray-to-Optical Lag in the Variations 
       of the Seyfert~1 Nucleus NGC~3516}

   \author{Dan Maoz\altaffilmark{1}, Rick Edelson\altaffilmark{2,3,4},
 \& Kirpal Nandra\altaffilmark{5,6}}

 \altaffiltext{1}{School of Physics \& Astronomy and Wise Observatory,
    Tel-Aviv University, Tel-Aviv 69978, Israel. dani@wise.tau.ac.il}

 \altaffiltext{2}{Astronomy Department, University of California, 
Los Angeles, CA 90095-1562}

  \altaffiltext{3}{X-ray Astronomy Group, Leicester University, Leicester LE1 7RH, 
United Kingdom}

   \altaffiltext{4}{Eureka Scientific, 2552 Delmar Ave., Oakland, CA 94602-3017}

    \altaffiltext{5}{NASA/Goddard Space Flight Center; Laboratory for High
Energy Astrophysics; Code 662; Greenbelt, MD 20771}

    \altaffiltext{6}{Universities Space Research Association}

\begin{abstract}

We present optical broadband ($B$ and $R$) observations of the Seyfert
1 nucleus NGC~3516, obtained at Wise Observatory from 1997 March to
1998 September, contemporaneously with the X-ray 2--10 keV
measurements of \rxte. The cross correlation function shows a positive
peak when the optical variations lead the X-rays by $\sim
100$ days, and anticorrelation peaks at various leads and delays
between the X-rays and the optical.
We show that the putative correlation
signal at 100 days 
is entirely due to the slow ($\gtorder 30$ days) components of
the light curves.  During the first year or this monitoring, smoothed
versions of the light curves are nearly identical copies of each other,
but scaled in amplitude and shifted in time. However, for the next 200
days, the X-ray and optical variations are clearly different. During
the whole period, the faster-changing components of the light curves
are uncorrelated at any lag.  

We consider the detection of these lags  tentative and the
significance of the correlations uncertain. If the 100-day delay
is real, however, one interpretation is that that the slowly-varying
part of the X-ray emission is an echo of the optical emission, Compton
scattered from a medium located at, or extending, $\sim 50-100$~light
days from the optical source.  We point out that a possibly analogous
phenomenon, of a lag between hard and soft X-rays for a given
variability timescale, exists in Galactic stellar-mass accretors.
Remarkably, in both cases the lag corresponds to a light travel
distance of order $10^4$ gravitational radii.  Alternatively, the lag
may not represent a physical size, but some other time scale. For
example, it may be the manifestation of an instability propagating
inward in an accretion flow, appearing first in the optical and then
in the X-rays. In any event,
we observe no strong correlation at zero lag, or at the small
positive lags expected if the optical continuum were
produced by reprocessing of X-rays. An energetically--significant 
reprocessed component in the optical emission of NGC 3516
is thus ruled out by our data.
 
\end{abstract}

\keywords{galaxies: active -- galaxies: individual (NGC~3516) --
galaxies: Seyfert -- x-rays: galaxies}

\section{Introduction}
Much of the energy of Seyfert-1-type
active galactic nuclei (AGNs) 
 is emitted in X-rays, yet it is unclear
what the source of this emission is.
Comparison of variations in different bands can 
provide valuable clues toward understanding
the geometry and nature  of AGNs.
In particular, inter-band lags can discriminate
between primary and secondary (i.e., reprocessed)
emissions. 

Contemporaneous X-ray and UV/optical monitoring has been carried out
for only a few Seyfert 1 galaxies to date.  On short time
scales, simultaneous optical and X-ray monitoring of both NGC~4051
(Done et al. 1990) and NGC~3516 (Edelson et al.  1999) showed strong
X-ray variations and little or no optical changes over 2--3~day
periods.  Longer time scale monitoring of NGC~5548 (Clavel et
al. 1992) and NGC~4151 (Kaspi et al. 1996; Crenshaw e al. 1996;
 Warwick et al.  1996; Edelson
et al. 1996) found evidence for a correlation at zero lag between optical,
ultraviolet ({\it IUE} data), and X-ray ({\it ROSAT} and {\it ASCA}
data), but these data were very sparsely sampled ($\le$12 points).
NGC~7469 was monitored intensively with \rxte, {\it IUE} and
ground-based observatories for one month in 1996.  The optical and UV
were found to be strongly correlated, with evidence presented for a
lag that increases with wavelength (Wanders et al. 1997; Collier et
al. 1998).  There was, however, no clear correlation found between the
X-rays and UV (Nandra et al. 1998).  The peaks in the X-ray light
curve appeared to lag the UV peaks by $\sim 4$~days, while the
troughs appeared better correlated at zero lag. The X-rays also showed
much more rapid variations than the UV and, by extension, the optical.
 Most recently, 
Chiang et al. (1999) monitored NGC~5548
for three days with \rxte, {\it ASCA}, and the {\it Extreme Ultraviolet
Explorer (EUVE)}. Evidence was presented for a lag that increases
with energy band, with the {\it ASCA} (0.5-1 keV) variations lagging
the {\it EUVE} (0.14--0.18 keV) variations by about 3.5 hours, and the 
\rxte (2--20 keV) variations lagging {\it EUVE} by about 10 hours.

We initiated in 1997 a program to monitor the Seyfert 1 galaxy
NGC~3516 with \rxte. Apart from its brightness and known tendency to
vary, the high declination of this galaxy makes it circumpolar for
most Northern ground-based observatories, allowing it to be observed year
round.  Month- and year-long variation timescales can thus be properly
probed, as well as shorter timescales.  Edelson \& Nandra (1999)
presented the \rxte~ data for NGC~3516 between 1997 March and 1998
September, and calculated the power-density spectrum (PDS) of the
2--10 keV fluctuations on all timescales from 20 min to 6 months. They
found that the PDS can be described by a power law of slope $-1.7$
that turns over to a flatter slope at timescales longer than $\sim 1$
month.

Here we present densely-sampled optical broad-band ($B$ and $R$) 
measurements of NGC~3516 obtained at Wise Observatory contemporaneously
with the \rxte~ monitoring, and supplement the \rxte~ light curve
with new data through 1999 January.
In \S2 we describe the observations and data reduction, and derive
the optical light curves. In \S 3 we carry out a time series analysis
comparing the X-ray and optical light curves. In \S 4 we attempt to
interpret our results within a physical picture.

\section{Optical Observations and Reductions}

We observed NGC~3516 from 1997, March 5, to 1998, September 2, using the
Wise Observatory 1m telescope in Mitzpe Ramon, Israel. On the
nights when the galaxy was observed, Johnson-Cousins
$B$- and $R$-band images were obtained once per night.
We  used a $1024\times 1024$-pixel
thinned Tektronix CCD at the Cassegrain focus, 
with a scale of $0.7''$~pixel$^{-1}$. Exposure times were 3 min in $R$
and 5 min in $B$.
During this 546-day period, useful data were obtained for 108 epochs
in $R$ and for 87 epochs in $B$. Between 1997 February 1 and November 25
the telescope suffered from scattered-light problems due to a change
in baffling. Data from this period could not be properly flat-fielded.
However, under proper baffling of scattered light one can see that
 the detector
response and illumination vary by only a few percent across the $12'$ field
of view of the detector, so there should only be
a minor effect on the accuracy of our photometry.
We verify this below.

Aperture photometry was carried out by integrating counts 
within circular apertures centered on the Seyfert nucleus and 
on the six brightest unsaturated stars projected near the galaxy.
The stars were chosen to be within a few arcminutes from the galaxy,
and in various directions, in order to minimize the error due to
the lack of proper flatfielding for most of the frames. On some
epochs, only the central section of the CCD was read out, and hence
not all six comparison stars are present on the frame.
Measurements of stars in which any of the pixels were near saturation
were discarded.
The apertures had a radius of 4 pixels. For comparison, the
seeing half-width at half-maximum (HWHM) was in the range of 1 to 2.5
pixels, with a typical value of 1.5 pixels. The aperture thus included
most of the light from a star, even under adverse seeing conditions.
 The local background level 
was calculated in annuli of inner and outer radii 8 and 11 pixels, respectively,
around each object.
For the measurement of the nucleus, this background subtraction
provides some removal of the galaxy starlight.
We experimented using smaller or larger apertures. We obtained
similar light curves for the nucleus,
but with smaller variation amplitudes for the larger apertures, due to
the larger constant stellar contribution. On the other hand, the errors
in the light curves (as determined below) also became larger for small
apertures, due to the dependence of the integrated counts on the 
object-centering accuracy in the pixellated images. We found that the 4-pixel 
aperture radius was optimal in terms of minimizing both the galaxy
background and the photometric errors.

Relative photometry was achieved by calculating the instrumental magnitude
difference between a star's counts in a given epoch and its counts in
 the first epoch of the program. These differences were averaged among all
the comparison stars present on a frame to provide an instrumental zero point
for a given epoch.
The standard deviation of this mean provided an empirical estimate
of the photometric error. The difference of the nuclear instrumental
magnitude and the zeropoint of a given epoch yielded the 
change in magnitude of the nucleus relative to the first epoch.
We verified that there is only a barely-discernible effect of the choice
of ``first epoch'' on the final light curves .

To assure that the comparison stars are not variable themselves,
and to assess the reliability of our error estimates, we 
measured in the same way each star using the five other stars as comparisons.
We found that the stars are non-variable to within our measurement
accuracy. The deviations of a star's brightness from its mean
are consistent with its assigned error-bars, assuming a Gaussian
error distribution. The mean error is 0.02~mag. For epochs whose
frames contained fewer than four comparison stars, the standard
deviation of the zeropoint mean was poorly defined, and the larger
among the standard deviation and 0.02~mag was adopted as the error.

Figure 1 shows the optical light curves we have obtained for NGC~3516.
In Figure 2 we plot on the same scale for each optical band the constant,
to within errors, 
 light curve of
one of the comparison stars, calculated relative to the other five stars.
The $R$ and $B$ light curves of NGC~3516 in Figure 1
 show very similar variability patterns,
with peak-to-peak amplitudes of 0.35 mag and 0.7 mag, respectively.
There is thus no doubt as to the reality of the variations. As
mentioned above, the exact amplitude of the variations depends
on the choice of photometric extraction aperture, which will include a particular
fraction of stellar light from the galaxy. The above numbers
are therefore lower limits on the {\it intrinsic} variability amplitude
of the nucleus in each band, which is difficult to estimate.

\section{Time Series Analysis}

Here we compare the X-ray and optical light curves of NGC~3516. 
All our 
results apply equally well to both the $B$ and the $R$
light curves, to which we will refer collectively as the ``optical
light curves''. Since the $R$ light curve is better sampled
than the $B$ light curve,
we will use only the $R$ in the figures and discussion below.
Figure 3 (top panel) shows again the $R$ light curve of NGC 3516,
but with a relative linear (rather than magnitude) flux scale.
The bottom panel shows
the \rxte~ X-ray (2-10 keV) light curve of Edelson \& Nandra (1999),
supplemented with new \rxte~ data up to January 1999. Observations and
reduction leading to the new \rxte~ data are as described in Edelson 
\& Nandra (1999).

Examination of Figure 3 shows that the bulk of the optical
variation is in a $\sim 250$-day-long rise and fall between
days 600 and 850, followed by a two-month-long deep
minimum centered around day 1000.
The X-ray light curve, by contrast,
has much more power in short-timescale flickering.

The z-transformed discrete correlation function (ZDCF;
Alexander 1997), a modification of the discrete correlation function
(Edelson \& Krolik 1988) was used to assess the degree of correlation
between variations in the optical and X-ray bands.
The top panel of Figure~4 shows the ZDCF for the unsmoothed $R$-band and 
X-ray data. A positive correlation of $ r = 0.70 $ is seen at a lag of 
$ \Delta t \approx -110 $~days (that is, with the optical variations 
leading the X-rays) while an anticorrelation of $r=-0.70$ is found 
at a lag of $ \Delta t \approx -280 $~days. Furthermore, anticorrelations
of $ r = -0.3 $ to $ r = -0.5 $
are seen between
lags of $ \Delta t \approx +100$ to $+200 $~days. 
We also note that there is a small subpeak close to zero lag.

The significance of the cross-correlation peaks is usually computed
using Student's t-test, with the null-hypothesis probability
depending on the number of independent points in the correlation. 
Usually, this is assumed to be the number of data points 
in each bin of the ZDCF, and under such an assumption 
the correlations we find are highly significant.
 Here we question
this assumption, however.
It is well-known that the
PDSs of AGN, including NGC~3516,
have a ``red-noise" character, with variations correlated over time
scales of $\le$1~month. The number of {\it independent} data points
in each correlation bin may therefore be greatly reduced, as will
the inferred significance of the correlations.
The high significance of both the positive and negative correlations 
also indicates that the underlying assumptions need to be examined.
With no straightforward way of estimating the number of independent
data points in a given bin, we caution that the
significances usually assumed are almost certainly overestimated.

To obtain a more quantitative assessment of the significance of the
correlation, we have carried out Monte Carlo simulations, as follows.
Synthetic light curves having chosen PDSs were created by summing
suitably-weighted harmonic functions with random phases. The synthetic light
curves were then sampled with the same temporal pattern as the real optical
and X-ray light curves. Simulated Gaussian measurement errors were
added as each point, such that the ratio of the rms variation of the
light curve to the Gaussian $\sigma$ was typical of that of the real
light curves. The cross-correlation function of the
simulated optical and X-ray light curves was then searched for values
as high as the one observed in the real data. The whole process was
repeated 1000 times for each choice of optical and X-ray PDS, and the
fraction of iterations with correlation above the threshold noted.

We find the results of these simulations are strongly dependent
on the assumed PDS of each light curve. Edelson \& Nandra (1999) showed
that the X-ray PDS of NGC~3516, on timescales shorter than about 1 day,
is well described by a power law of index $\alpha_x=-1.74\pm 0.12$. On longer
timescales, however, the index gradually flattens, to  $\alpha_x\approx -1.0$
on day-long  to month-long timescales, and further to $\alpha_x\approx -0.7$
on few-month timescales. The PDS slope on timescales longer than one day,
precisely the timescales probed here, is not well contrained. The observational
knowledge of the optical PDS is much worse. The uneven sampling of the
optical light curve precludes any straightforward calculation of its PDS.
Existing algorithms, e.g. Scargle (1982), for calculating the PDS
of unevenly sampled data, are useful for periodicity searches, but badly
fail to reproduce the shape of PDS's having power over a broad range
of frequencies, due to the aliasing between frequencies that the uneven
window function introduces. (See Giveon et al. 1999,
for a detailed discussion of the problem.) To obtain a very rough guide
of the optical PDS shape, we applied to the data Giveon et al.'s (1999) 
``partial interpolation'' algorithm. The results suggest the optical PDS may
be a power law of slope $\alpha_o\sim -2.0\pm0.6$.

Given the above uncertainties with regard to the PDS's that must be input
to the simulations, we calculated the significance of the observed
correlation for a grid of power-law PDS's with different slopes.
The significance of the correlation is highest for flat input PDS slopes,
and becomes low for steep PDS's, in which each light curve is
dominated by only a few ``events'' which can produce spurious correlations.
For an input X-ray PDS slope of $\alpha_x=-1.0$, which is a reasonable choice,
the observed correlation is significant at $>99\%$ confidence,
 as long as the optical PDS slope $\alpha_o\ge -1.5$. For $\alpha_o=-1.75$
the significance declines to $98.5\%$, and for $\alpha_o=-2.5$ it is only $97\%$.
Steeper optical PDS's are allowed for $\alpha_x$ somewhat flatter than $-1.0$,
and vice versa. We conclude that the observed X-ray-to-optical correlation 
at 110-day lag may indeed be significant, but the verdict depends on poorly known
parameters. We also note that, as shown below, the observed correlation is actually
driven only by the first year's worth of data, during which the correlation
is much higher. However, calculating the significance of only a segment of the data
obtained would involve {\it a posteriori} statistics, which is something
we will avoid.

To study the relative contributions to the correlations
made by fast and slow variations, we have smoothed the light
curves with a 30-day boxcar running mean, and recalculated the
ZDCF. The smoothed light curves are shown as solid lines in Figure 3.
The middle panel of Figure~4 shows the ZDCF for the smoothed data, and
the bottom panel, the ZDCF for the residuals (i.e., the original
light curves minus their respective smoothed versions).
The smoothed light curves show 
 correlations and anticorrelations that are similar, 
but somewhat strengthened, compared to the unsmoothed
light curves, with a positive peak of $ r = 0.80 $ at 
$ \Delta t = -100 $~days, a negative peak of $ r = -0.90 $ at 
$ \Delta t = -280 $~days, and a negative plateau of $ r \approx -0.65 $ at
$ \Delta t = +100$ to $+200 $~days. 
The correlation function of the residual light curves
shows no significant signal,
indicating that there is no correlation present in the high temporal
frequency components of the data.

We obtain similar results if, instead of using the ZDCF algorithm,
we use a ``least-squares shift and scale'' scheme to
find the best lag for the observed, smoothed, or residual light curves.
For every time-shift between the light curves, we find the linear relation
that, when applied to the X-ray light curve,
 minimizes the sum of the square of the differences between each
 optical point and the scaled X-ray point that is nearest in time to it
at that shift. The global (over all time shifts) least squares
then provides the best lag.

On the face of it, the positive peak in the correlation suggests that the
optical variations lead the X-ray variations by $\sim 100$d.
The similarity between the
smoothed optical and X-ray light curves during the first 
350 days can be seen in Figure 5,
which shows the two after the X-rays are scaled, and shifted
back in time, according to the least-squares solution for this period.
This figure clarifies the fact that 
the strong apparent correlation we observe is driven primarily
by one ``event'' in the light curve. Indeed, one sees a complete mismatch,
at this lag, of the light curves after day 850. The deep minimum
observed in the optical light curve seems to correspond at zero
lag to a local minimum in the X-ray light curve, which may account for
the subpeak in the ZDCF at zero lag. This increases the likelihood
that we are perhaps being misled by chance similarities in temporal
structure of parts of the light curves, while in fact
there is no correlation between the variations in the different bands.
Also, the anti-correlations
mentioned above have similar significance to that of the positive peak, and
while there have been no physical mechanisms proposed to explain an
anticorrelation between the bands, there is no statistical reason to
prefer positive correlations over negative ones. 
Continued monitoring may help discriminate between these options,
whose physical implications we discuss in the next section.

We have also cross-correlated
the optical light curves themselves. The ZDCF of the 
$B$ vs. $R$ light curves (Figure 6) shows, as expected, that they are
highly correlated ($r=0.95$), but the peak and the centroid of the ZDCF
are slightly shifted from zero lag, indicating the $R$ variations lag
the $B$ variations by several days (which is comparable to the mean
sampling interval).
While this delay could be interpreted as a wavelength-dependent
continuum lag, we believe a more likely explanation
is the fact that the strong broad H$\alpha$ line is included in the 
$R$ band, and contributes of the order of 10-20\% of the broad-band
flux. Balmer-line variations in this galaxy lag the
continuum variations by about 11 days (Wanders et al. 1993) due to the
light-travel time across the broad-line region.
The H$\alpha$ contribution to the $R$ band probably shifts slightly  the
ZDCF peak from the peak at or near zero lag that it would have
if there were only variable continuum emission in the band.
The observed delay may therefore be considered an upper limit
on the true delay between $B$ and $R$.

\section{Discussion}

Much current thinking about the emission processes in AGNs
centers around the notion that the X-rays arise from very close
(within a few 
Schwarzschild radii, $R_S$) of
a massive black hole. Support for this idea
has come from the rapid variability that is observed in
X-rays (implying small physical scales), as well as the
detection in X-rays of a broad Fe K-shell emission line
in many Seyfert 1s (e.g. Nandra et al. 1997). The emission line is
thought to be gravitationally and Doppler
broadened fluorescence of the inner parts of an accretion disk, after
the disk is illuminated by the X-rays.
The continuum-emission mechanism is not
known, but most commonly it is assumed that the X-rays are optical/UV photons
which have been upscattered by a population of hot electrons.
The acceleration mechanism and geometry of the X-ray source is not
known. Neither is the source of seed photons, and despite some
substantial problems it is still usually assumed that the optical/UV
arises directly from an accretion disk (Shields 1978; Malkan 1983).
It has also been hypothesized that X--rays illuminating the disk,
or other optically thick gas, might be responsible for some or all
of the optical/UV radiation, via reprocessing (Guilbert \& Rees 1988;
Clavel et al. 1992).

Variability data
such as those we have presented above can provide stringent constraints
on possible models. 
In summary, our data have shown strong variability in both optical and
X-ray bands on month-to-year time scales, but rapid (days) variations 
only in the X-rays. The zero-lag correlation between the bands is poor,
with a much stronger relationship implied if the optical variations
lead those in the X-rays by $\sim 100 d$. 
This ``100-day lead'' breaks down in the latter parts of the monitoring period.
Strong negative correlations are also 
observed for optical leads of $280$ d and
 optical {\it lags} of $100-200 d$. This makes us cautious about the
reality of the lag, but we will discuss some physical
implications of our results below.

It has long been suspected that the X-rays show more rapid variations
than the optical/UV, and this has been explicitly demonstrated in a few
cases (e.g., NGC~4051, Done et al. 1990; NGC~7469, Nandra et al. 1998;
NGC~3516, in a 3-day  HST/RXTE/ASCA campaign, Edelson et al. 1999). 
Our data add to that
body of evidence, which implies, unavoidably, either that the 
observed optical radiation is
not the primary seed photon source, or that the process which turns
these photons into X-rays induces variability intrinsically. Given a supposed
location in the inner few $R_{\rm S}$, it might be more natural to
assume that UV or EUV photons are the seeds for the X-rays. With
an origin in the inner disk, the EUV emission would be expected to
be more variable, although it is still extremely difficult to reconcile
variability as rapid as that observed with physical time scales in 
the disk (Molendi, Maraschi \& Stella 1992). The X-ray source
itself
may be less directly connected to the disk physics, and could,  in principle,
change much more rapidly, especially if consisting of multiple flaring
regions, as opposed to a single, coherent one.

We have also shown tentatively that the X-rays may respond
on long ($\sim 100d$) time scales to variations in the optical.
One way of viewing the 2--10 keV emission, then, 
is as the sum of two components: 
a smoothly-varying component
that is very similar to the optical light curves during the first year, 
but lags them by $\sim 100$ days, and a fast, flickering, component that is
uncorrelated with the optical variations.
The seeming lack of a deep minimum in
the X-ray light curve, corresponding to a delayed version of the
minimum seen in the optical light curve around day 1000, 
could arise because the slow, delayed component had nearly
turned off, and the X-ray emission had become  dominated by the second
component. Indeed, if a constant is subtracted from the smoothed
X-ray light curve, such that the smoothed curve always passes below 
the observed 
X-ray measurements (to ensure that the flux in the fast component
is always positive), then the lowest points in the smoothed
light curve just reach zero flux (see Figure 3).
The delay of the smooth component behind the optical emission
could then be interpreted as the light-travel time between
the seed photon source and a Compton upscattering region.
If it is not the optical 
photons themselves being upscattered, we would need to assume that the
 variations of the optical light curves can serve as surrogates
for some other seed photon (UV or soft X-ray) variations.
The large delay observed
would put the scatterer at a relatively large distance, 
$r\approx 50-100$~lt-days ($1.25-2.5\times10^{17}$cm) from the nucleus, i.e. 
$\sim 10^4 R_S$
(for a $10^8 M_{\odot}$ black hole).

It is possible to imagine a toy model
of the required ``Compton mirror'' by surrounding the nucleus
with a $T\approx 3\times10^9 K$ (or hotter, 
depending on the energy of the seed photons) electron gas
in a thin ($\Delta r\ltorder 20$ lt-days $=5\times10^{16}$cm)
50-lt-day-radius shell of density $n\approx2\times10^6$ cm$^{-3}$
and column density $N\approx 10^{23}$ cm$^{-2}$, giving a low optical
depth to Compton scattering of $\tau\sim 0.1$. The geometrical 
thinness is contrained by the small amount of broadening allowed
by the data between the X-ray and optical pulse.
This configuration would ensure that only photons that are singly
backscattered by large angles
are upscattered to the 2-10 keV band, so that a coherent echo
is seen only from a small cap on the far side of the shell. 
This optical depth
will also roughly produce the observed ratio of 1 keV and 3 keV
photons in this object, corresponding 
to the photon index $\Gamma\approx 2$ between 0.6--10 keV 
measured by George et al. (1998). However, more sophisticated
calculations are required to see if the observed variations
and detailed spectrum can be reproduced in this scenario.
As already mentioned, this picture also leaves the rapid variability of the
X-rays unexplained, so we must then invoke either another X-ray
producing-region closer to the black hole, or an extended
region, which produces emission at various radii (and therefore
variations with a range of time scales).

Similar models have been proposed to explain the emission 
in Galactic stellar-mass accreting neutron stars
and black holes. For individual bins of frequencies in the X-ray Fourier
spectrum of such objects,
the hard X-rays lag the soft X-rays by an approximately-constant
phase, meaning there is a time lag that increases linearly 
with the Fourier timescale probed.  For low frequencies (0.1 Hz)
the time lag is about  0.2 s, corresponding, 
again, to $10^4 R_S/c$ (e.g., van der Klis et al . 1987; 
Miyamoto et al. 1992; Vaughan et al. 1994;  Ford et al. 1999).
Here, too, it has been proposed that the lags are due to light-travel
time in a very extended 
Compton upscattering gas (e.g., Sunyaev \& Truemper 1979;
Payne 1980; Kazanas, Hua, \& Titarchuk 1997; Hua, Kazanas \& Cui 1999). 
The more recent of these
models invoke a centrally-concentrated distribution of gas, 
although the similarity of the optical and X-ray light curves in
NGC~3516 are suggestive of the thinner shell referred to above.

This picture is not free of problems.  A separation into two (or more)
X-ray components may be considered {\it ad hoc}. There are also
strong implications from the spectral observations that the 
bulk of the X-ray continuum is concentrated in the central regions,
arguing against a region extending to many thousand gravitational radii.
We have not considered how the Compton-upscattering gas is 
heated to its high temperature at such a large radius. 
A similar
problem is encountered for the X-ray binaries (e.g. Stollman et al. 1987).
Solutions that have been suggested include that, for black holes,
the gas is heated locally as part of an advection dominated
accretion flow (ADAF, e.g. Narayan \& Yi 1994), or, for neutron stars, that the
gas was preheated by radiation from the central source (Kazanas et al. 1997),
or that the energy is transported via magnetic fields (Stone et al. 1996).

Alternatively, the $100 d$ signal may be associated not with
a light-travel time, but with some other time scale.
One interpretation that is more in line
with standard thinking about the inner regions of AGNs
is that we are witnessing the
effects of an inhomogeneous accretion flow onto the
black hole. The time
inferred from the optical-X-ray lag 
is then some timescale associated
with the disk, or accretion, process.
Some form of
instability, e.g., thermal or viscous, forms in the disk and
causes its optical emission
to brighten. The instability then propagates inwards
to the hotter, X-ray emitting radii, on
a timescale of 100 days, when an X-ray ``copy'' of it is seen in
the light curve. One problem with this scenario, however, is 
that it is unclear why the shapes of the variability 
patterns would be so similar in the two bands during the first year, 
implying that
the instability spent very nearly equal time intervals
in the optically-emitting and X-ray-emitting regions.
The lack of correlation in the latter parts of the observation
is also unexplained, meaning that the
processes in the inner disk are far more complex than what we have 
just described.

These kind of explanations have also been put forward
for the hard X-ray lags in Galactic accretors. For example,
Orosz et al. (1997) found that the optical brightening
of the ``microquasar'' GRO-J1655-40 preceded its X-ray (2-12 keV)
outburst by 6 days, and suggested this was the result of an inward
propagation of a disturbance in the accretion disk.
Bottcher \& Liang (1999) also presented models 
for accretion of a cool blob in an advection dominated flow,
with the blob's radiation being Compton upscattered by a progressively 
hotter and denser corona as it drifts toward the event horizon at
constant radial velocity. Alternatively,
Poutanen \& Fabian (1999) suggest that the lags reflect the timescale
for development of a magnetic flare that floats out of a thin accretion
disk into a hot, optically-thin, corona, emitting 
progressively harder radiation until the flare ends suddenly. 
Applied to AGNs, their model has the attraction of naturally 
maintaining consistency
with the fast variability time scales in the X-ray, as well as the
strength and extreme broadening of the iron K$\alpha$ line.
These pose serious problems for both the extended-corona and the ADAF
models.

We must also keep in mind the possibility that the similarity
of the X-ray and optical 
light curves at 100-days lag during the first year
of our program may just be a chance coincidence,
and that this is the reason for the extreme differences 
between the light curves after the first year.
Even if there is no real correlation between
the variability in these different bands,
our data still constrain the origin of the optical emission.
We observe no strong correlation at zero lag, or at the small
positive lags expected if the optical continuum were
produced by reprocessing of X-rays. An energetically--significant 
reprocessed component in the optical emission of NGC 3516
is ruled out by our data (c.f. NGC~7469, Nandra et al. 1998).

Interpretation aside, 
we also note that both the 100-day lag between X-rays and optical,
and the 30-day timescale that separates
the slow, possibly-correlated X-ray variations from the fast, uncorrelated
X-ray flickering, are similar to the turnover timescale in the PDS
found for this object by Edelson \& Nandra (1999). It will be
interesting in the future to construct more specific models
which can tie together these time scales.

\acknowledgements

We would like the thank the following observers at Wise,
who contributed their efforts and observing time to obtain
the data presented here:  R. Be'eri, T. Contini,
J. Dann, A. Gal-Yam, U. Giveon, A. Heller, S. Kaspi, Y. Lipkin,  I. Maor,
H. Mendelson, E. Ofek,  A. Retter, O. Shemmer,
 G. Raviv, and S. Steindling. We are also grateful for the assistance
of the Wise Observatory staff: S. Ben-Guigui, P. Ibbetson, and E. Mashal.
We acknowledge valuable discussions with N. Arav, I. George, A. Laor,
A. Levinson, H. Netzer, A. Sternberg, and J. Turner. T. Alexander is
thanked for providing his ZDCF code, and  the anonymous referee
for useful suggestions.
Multiwavelength studies at Wise Observatory
are supported by a grant from the Israel Science Foundation.

\begin{figure}
\epsscale{1.1}
\plotone{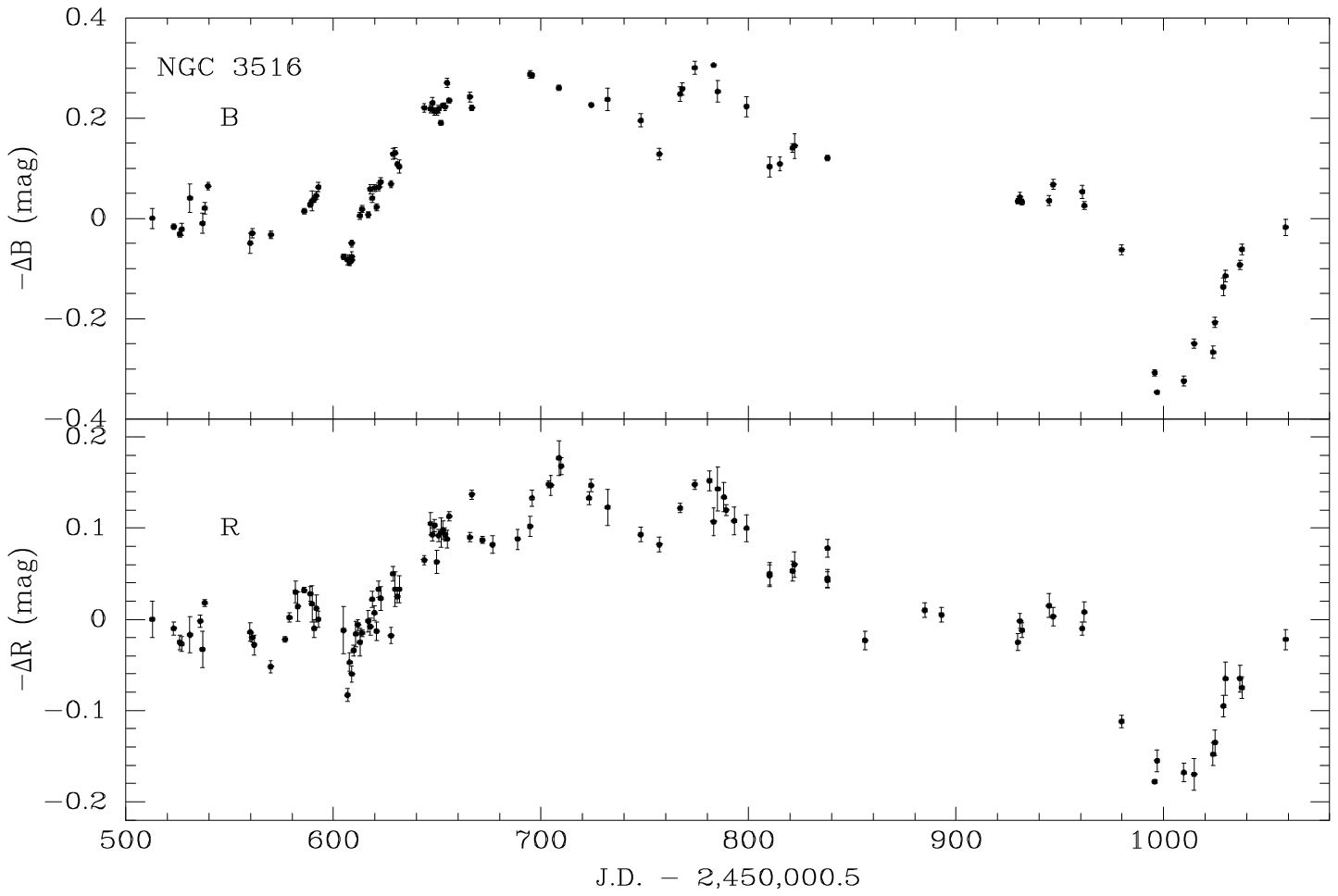}
\caption{$B$-band (top panel) and $R$-band (bottom panel)
 light curves for NGC~3516. 
}
\end{figure}

\begin{figure}
\epsscale{1}
\plotone{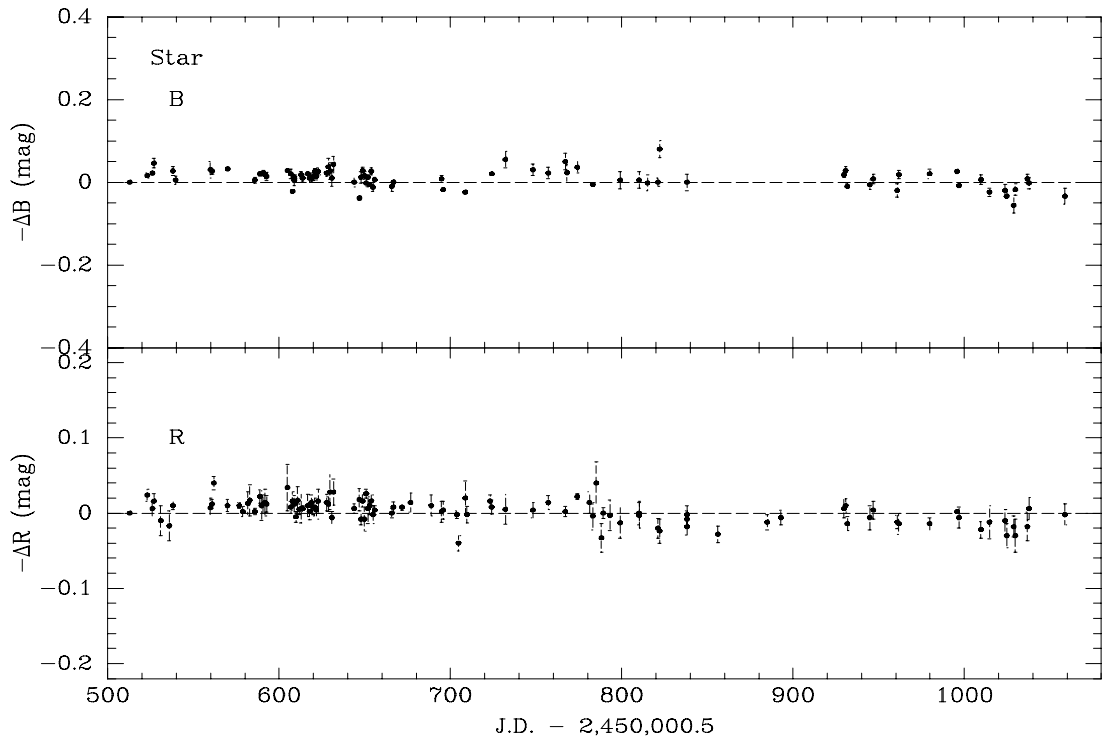}
\caption{$B$-band (top panel) and $R$-band (bottom panel)
 light curves for one of the comparison stars, measured in the same way
as the Seyfert nucleus. 
}
\end{figure}

\begin{figure}
\epsscale{1}
\plotone{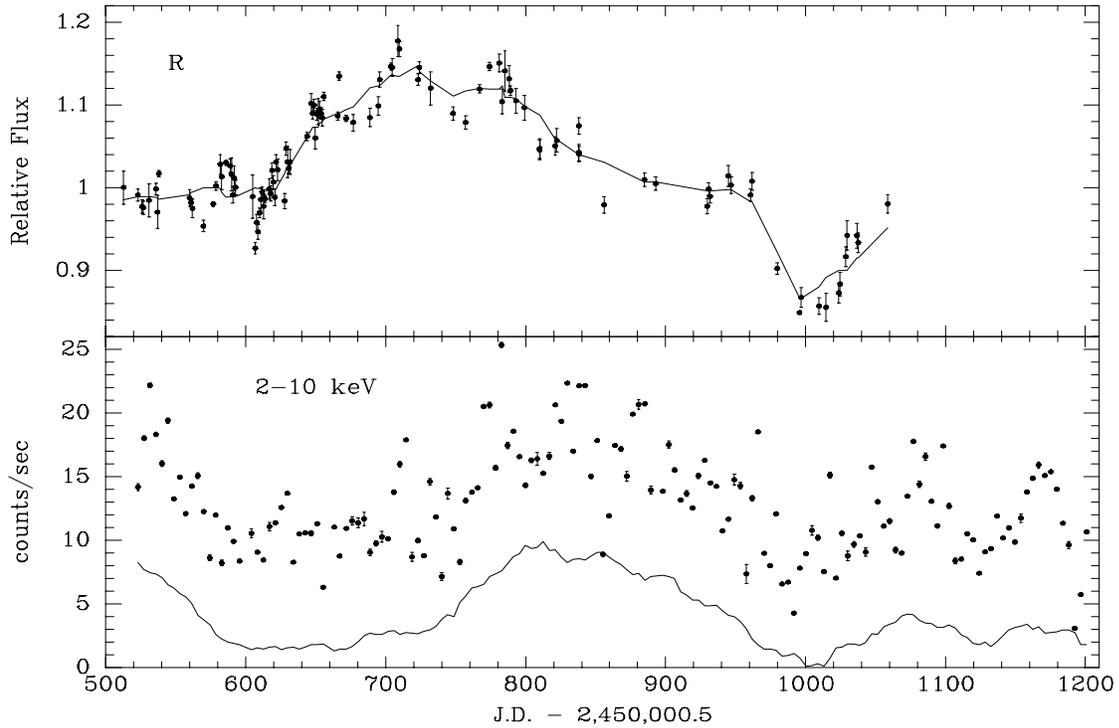}
\caption{Top panel: $R$-band light curve of NGC 3516,
but with linear flux scale. Solid line is 30-day boxcar 
smoothed version of the light curve.
Bottom panel:
 \rxte~ X-ray (2-10 keV) light curve of Edelson \& Nandra (1998),
supplemented with new \rxte~ data up to January 1999. 
Solid line is a 30-day smoothed version of the X-ray light curve,
vertically shifted so as to lie below all the X-ray measurements.
}
\end{figure}

\begin{figure}
\epsscale{1}
\plotone{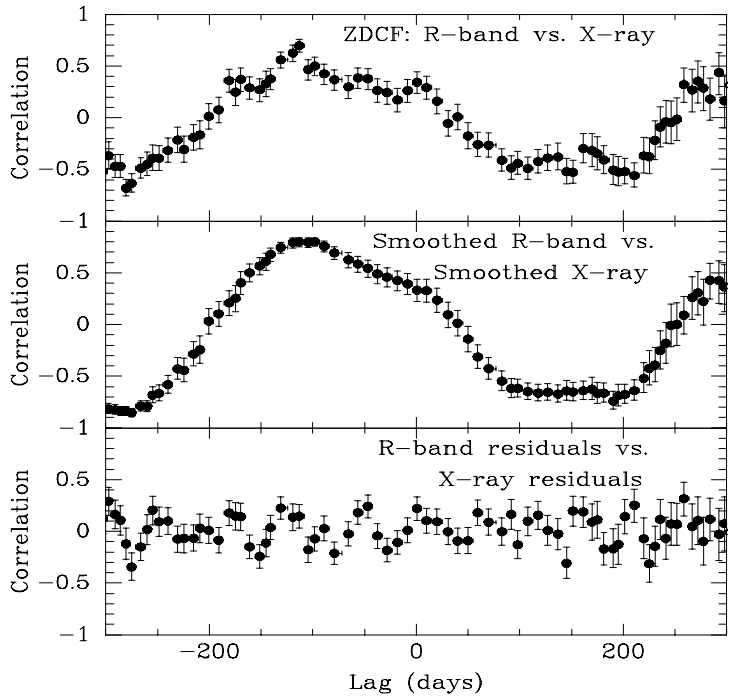}
\caption{ZDCF cross-correlations. Top panel: $R$-band vs. X-ray light 
curves. 
Middle panel: Same as above, but after the light curves have
been smoothed with a 30-day boxcar running mean (see Fig. 3). 
Bottom panel: As above, but between the
fast components of the X-ray and $R$-band light curves,
which are obtained by subtracting from each light curve
its smoothed version.}
\end{figure}

\begin{figure}
\epsscale{1}
\plotone{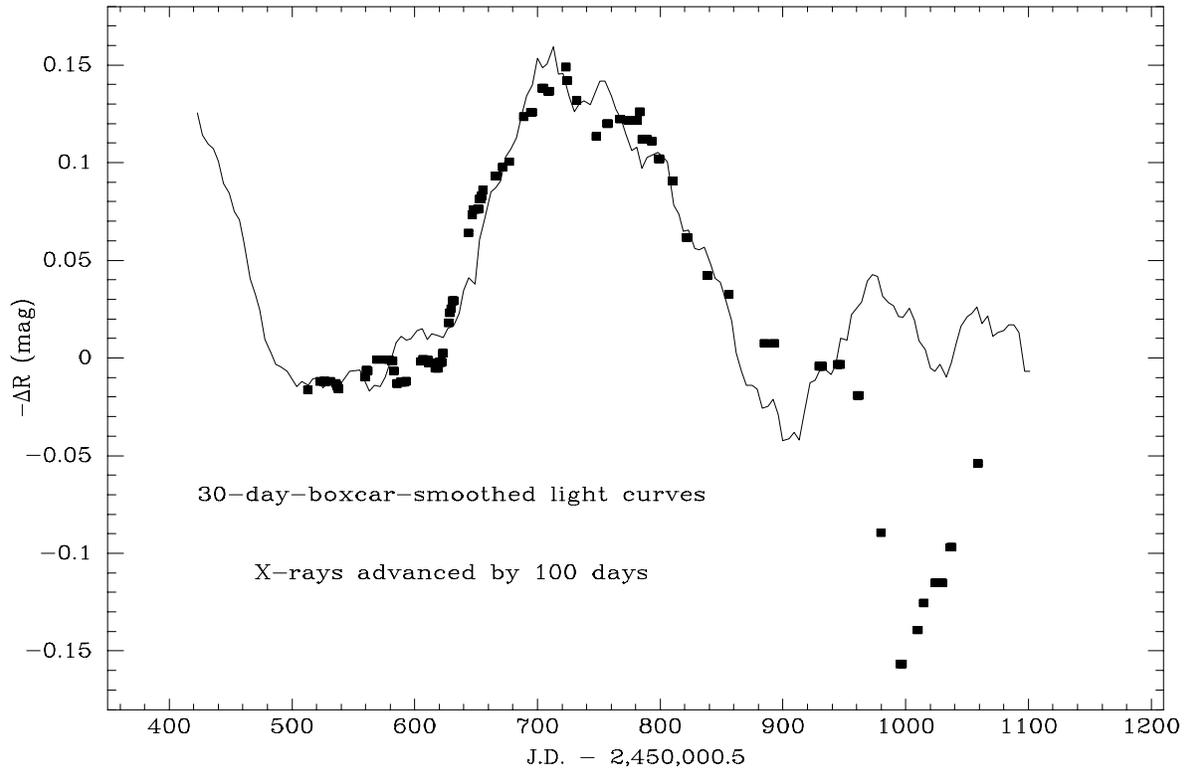}
\caption{Smoothed $R$-band light curve (boxes), and
smoothed X-ray light curve (solid line) after the latter
was advanced by 100 days and scaled linearly to obtain
the best match between the two during the
first year of data. Note the similarity
of the two smoothed light curves during the
first year, and the dissimilarity thereafter.}
\end{figure}

\begin{figure}
\epsscale{1.2}
\plotone{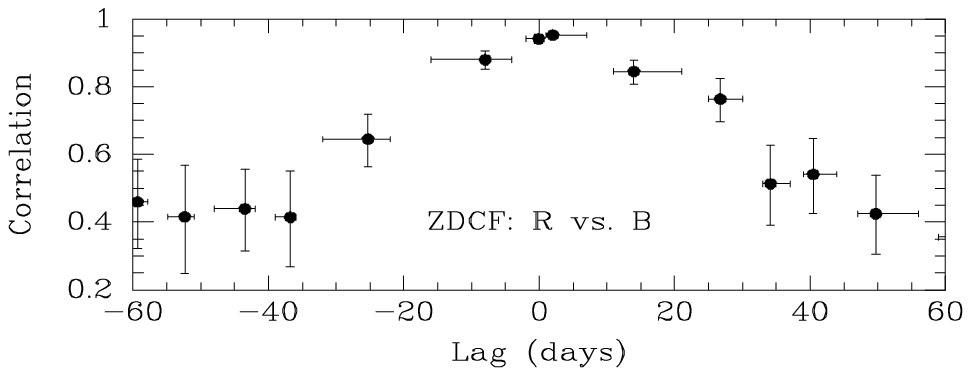}
\caption{ZDCF cross-correlation between the $B$ and
$R$ optical light curves. The small lag of $R$ behind
$B$ which is implied is likely due to the broad H$\alpha$ 
contribution to the $R$ band.}
\end{figure}


\begin{references}
\reference{} Alexander, T. 1997, in ``Astronomical Time Series'', eds. D. Maoz, A. Sternberg,
  and E.M. Leibowitz (Dordrecht: Kluwer), 163
\reference{} Bottcher, M., \& Liang, E.P. 1999, ApJL, in press
\reference{} Chiang, J., et al. 1999, ApJ, submitted
\reference{} Clavel, J., et al. 1992, ApJ, 393, 113
\reference{} Collier, S.J. et al. 1998, ApJ, 500, 162
\reference{} Crenshaw, D.M. et al. 1996, ApJ, 470, 322
\reference{} Done, C. et al. 1990, MNRAS, 243, 713
\reference{} Edelson, R.A., \& Krolik, J.H. 1988, ApJ, 333, 646
\reference{} Edelson, R.A.,  et al. 1996, ApJ, 470, 364
\reference{} Edelson, R., \& Nandra, K. 1999, ApJ, 514, 682
\reference{} Edelson, R. et al. 1999, in preparation
\reference{} Ford, E.C., van der Klis, M., Mendez, M., van Paradijs, J., \& Kaaret, P.
  1999, ApJ, 512, L31
\reference{} George, I., Turner, T.J., Netzer, H., Nandra, K., Mushotzky, R., Yaqoob, T. 1998,
  ApJS, 114, 73
\reference{} Giveon, U., Maoz, D., Kaspi, S., Netzer, H., \& Smith, P.S. 1999, MNRAS, 306, 637
\reference{} Guilbert, P.W.,  \& Rees, M.J. 1988, MNRAS, 233, 475
\reference{} Kaspi, S., et al. 1996, ApJ, 470, 336
\reference{} Kazanas, D., Hua, X.-M. \& Titarchuk, L. 1997, ApJ, 480, 735
\reference{} Malkan, M.A. 1983, ApJ,  268, 582
\reference{} Miyamoto, S., et al. 1992, Ap, 391, L21
\reference{} Molendi, S.,  Maraschi, L., \& Stella, L. 1992, MNRAS, 255, 27
\reference{} Nandra, K., George, I.M., Mushotzky, R.F., Turner, T.J., 
\& Yaqoob, T. 1997, ApJ, 477, 602
\reference{} Nandra K., et al. 1998, ApJ, 505, 594
\reference{} Narayan, R. \& Yi, I. 1994, ApJ, 428, L13
\reference{} Orosz, J.A., Remillard, R.A., Bailyn, C.D., \& 
McClintock, J.E. 1997, ApJ, 478, L83
\reference{} Payne, D.G. 1980, ApJ, 237, 951
\reference{} Poutanen, J. \& Fabian, A.C. 1999, MNRAS, 306, L31
\reference{} Scargle, J.D. 1982, ApJ, 263, 835
\reference{} Shields, G.A. 1978, Nature, 272, 706
\reference{} Stollman,G.M. et al. 1987, MNRAS, 227, 7P
\reference{} Stone, J.M., Hawley, J.F., Gammie, C.F., \& Balbus, S.A. 1996, ApJ, 463, 656
\reference{} Sunyaev, R.A.,  \& Truemper, J.  1979, Nature, 279, 506
\reference{} van der Klis, M., et al. 1987, ApJ, 319, L13
\reference{} Vaughan, B., et al. 1994, ApJ, 421, 738
\reference{} Wanders, I., et al. 1993, A\&A, 269, 39
\reference{} Wanders, I., et al. 1997, ApJS, 113, 69
\reference{} Wang, J.X., Zhou, Y.Y., Xu, H.G., and Wang, T.G 1999, ApJ, 516, L65
\reference{} Warwick, R.S.,  et al. 1996, ApJ, 470, 349
\end{references}
\end{document}